\documentclass[brazilian,twocolumn,aps,prb,superscriptaddress,amsmath,assume,showpieces,floatfix,preprintnumbers,longbibliography]{revtex4-2}
\usepackage[latin9,utf8]{inputenc}
\usepackage{float}
\usepackage{textcomp}
\usepackage{amsmath,amssymb,amscd,hyperref}
\usepackage{graphicx}
\makeatletter
\usepackage{url}
\usepackage{grffile}
\usepackage{bbold}
\usepackage{comment}
\usepackage{dirtytalk}
\usepackage{chngcntr}
\usepackage{tabularx}
\usepackage{multirow}
\usepackage{soul}
\usepackage{cleveref}
\usepackage{braket}
\DeclareUnicodeCharacter{2212}{-}
\DeclareUnicodeCharacter{0301}{\hspace{-1ex}\'{ }}

\usepackage{dcolumn}
\usepackage{color}
\DeclareGraphicsExtensions{.png .jpg .pdf}
\usepackage{hyperref}
\hypersetup{
     colorlinks = true,
     linkcolor = blue,
     anchorcolor = blue,
     citecolor = blue,
     filecolor = blue,
     urlcolor = blue
     }
\usepackage{braket}
\usepackage{physics}
\usepackage{array}
\usepackage{booktabs}
\usepackage{soul}
\makeatother

\begin{document}

\title{Magnetic field-tuned size and dual annihilation pathways of chiral magnetic bobbers}

\author{S. Y. Lu}
\affiliation{School of Materials and Physics, China University of Mining and Technology, Xuzhou 221116, P. R. China}
\affiliation{Key Laboratory of Magnetism and Magnetic Functional Materials of MoE, Lanzhou University, Lanzhou 730000, China}

\author{Y. F. Duan}
\email{yifeng@cumt.edu.cn}
\affiliation{School of Materials and Physics, China University of Mining and Technology, Xuzhou 221116, P. R. China}

\author{D. X. Yu}
\email{yudx@lzu.edu.cn}
\affiliation{Key Laboratory of Magnetism and Magnetic Functional Materials of MoE, Lanzhou University, Lanzhou 730000, China}

\author{H. M. Dong}
\email{hmdong@cumt.edu.cn}
\affiliation{School of Materials and Physics, China University of Mining and Technology, Xuzhou 221116, P. R. China}

\date{\today}

\begin{abstract} 
Magnetic chiral bobbers (CBs) are three-dimensional (3D) topological spin textures that consist of a tapered skyrmion tube terminating in a Bloch point, promising applications in high-density spintronics. However, the mechanisms controlling their size and the dynamics of their annihilation are still not fully understood. In this study, we present an analytical model that predicts the radius $R$ of the CB as a function of the external magnetic field, the Dzyaloshinskii-Moriya interaction (DMI), the magnetic anisotropy, and the exchange interaction. The micromagnetic simulations validate this model across a broad range of parameters. We also identify two mechanisms of annihilation of CBs: (i) a droplet-like instability that occurs under rapid changes in the magnetic field, which we describe using a proposed magnetic Weber number $We$ and its critical field step scaling; and (ii) Bloch point depinning mechanism at interfaces, for which we determine the threshold magnetic field $B_{\text{th}}$ for annihilation. Importantly, we uncover a novel fragmentation pathway in which CBs transform into skyrmion tubes, then into half-CBs, and finally into ferromagnetic states. These findings lay the groundwork for understanding and manipulating 3D CBs as next-generation devices.
\end{abstract}
\maketitle

\section{Introduction} 
Nontrivial magnetic topological structures, such as skyrmions, have garnered significant attention due to their topologically protected properties and potential in next-generation spintronic devices \cite{reichhard2022, zhang2023, cui2023}. In two-dimensional (2D) systems, the mechanisms of size tuning, creation, and annihilation of skyrmions have been extensively studied and experimentally validated \cite{wuSize2021, repicky2021, NatPh2022}. However, 3D magnetic topological textures remain comparatively unexplored, primarily due to their theoretical complexity and experimental challenges \cite{donnelly2022, pathakT2021}. Currently, the manipulation of 3D magnetic topological structures is quite limited. The intricate spatial magnetization distributions in 3D systems introduce additional degrees of freedom and energy contributions, making their stabilization and manipulation fundamentally distinct from 2D counterparts \cite{donnelly2017, zhengHopfion2023}. Experimentally, in-depth analysis of the internal structure of 3D magnetism remains a current technical bottleneck \cite{repicky2021}. In contrast, theoretical analysis and simulations can break through the limitations of experimental conditions and provide detailed theoretical support for experimental research \cite{Mich2018, liuSMonte2007}. Advancing the understanding of 3D topological states is crucial for expanding the horizon of magnetic soliton-based technologies, including ultra-dense memory and neuromorphic computing architectures \cite{ranaThree2023, ranBobber2021}.  

Among 3D topological structures, the magnetic chiral bobber (CB), a hybrid entity comprising a tapered skyrmion tube terminating in a Bloch point, has emerged as a promising candidate for high-density storage \cite{rybakov2015}. This new chiral quasiparticle opens up new opportunities for exploring intricate 3D magnetic topological structures both theoretically and experimentally \cite{rybakovNew2016, zhengExper2018}. It is experimentally found that the skyrmion phase can transform to chiral bobbers (CBs) by tuning the FeGe film thickness \cite{ahmedChiral2018, zhu2021}. A CB lattice is created in helimagnet-multilayer heterostructures via the proximity of two skyrmion states with comparable size \cite{ranBobber2021}. Moreover, studies have shown that CBs possess unique transport fingerprints that facilitate precise electrical detection \cite{redies2019}. The Hall angle of a CB can be precisely adjusted across a wide range, spanning from $−90^{\circ}$ to positive angles \cite{gong2021}. Recently, it has been shown that coupled 3D CBs exhibit a unique fingerprint that can be used to identify complex 3D magnetic textures via their dispersion curves \cite{bassir2022}.

While prior research has established its formation conditions and static stability, critical aspects remain unresolved. Unlike skyrmions, the size-tuning mechanisms of CBs under external fields and material parameters, as well as their annihilation dynamics, are poorly understood, which is a significant difference from these very different skyrmions. This knowledge gap impedes the development of CB-based functional devices, such as energy-efficient racetracks or topological transducers \cite{ahmedChiral2018, zhu2021}. Furthermore, the interplay between the CB's 3D geometry and its response to magnetic fields introduces unique physical phenomena absent in lower dimensions, necessitating dedicated theoretical investigation \cite{qiuSpin2024}.  

In this work, we address these challenges by establishing a comprehensive 3D micromagnetic model to unravel the size modulation and dual annihilation pathways of CBs. First, we derive an analytical expression for the CB radius $R$ as a function of the external magnetic field, the chiral Dzyaloshinskii-Moriya interaction (DMI), the anisotropy, and the exchange interaction. Second, we identify two distinct annihilation mechanisms: (i) a droplet-like instability under rapid field changes governed by a magnetic Weber number $We$, and (ii) Bloch-point depinning at interfaces dictated by anisotropy gradients and chiral DMI. Micromagnetic simulations validate our model across a range of parameters, confirming the scaling laws and the annihilation magnetic-field threshold for dynamic annihilation. Interestingly, we reveal a novel annihilation pathway via fragmentation of skyrmion tubes, unique to 3D topology. These findings can provide foundational principles for understanding and manipulating CBs in future spintronic applications.  

\section{Theoretical model} 
\subsection{Size of CBs}

Theoretical and experimental studies have demonstrated that CBs can exist in stable states \cite{ahmedChiral2018, zhu2021, ranBobber2021, redies2019}. In this work, we establish a 3D magnetic topological model to investigate the size and annihilation mechanisms of CBs, combining micromagnetic theoretical and analytical methods with numerical simulations. CBs are quasi-3D magnetic structures composed of skyrmions with radii $R$ that gradually decrease with the depth $z$ increasing. The radius $R$ of surface skyrmions can be measured experimentally \cite{dong2023}. $h$ denotes the height of the CBs, as illustrated in Fig. 1(a). For each layer of the skyrmion in a CB, we consider components along the axial direction $\mathbf{e}_z$ and the radial direction $\mathbf{e}_r$. The unit magnetization vector distribution in cylindrical coordinates $(r, z)$ can be written as
\begin{equation}
\mathbf{m} =  \sin\theta \mathbf{e}_r+ \cos\theta \mathbf{e}_z,   
\end{equation}
where $\theta$ represents the angle between the magnetization vector and the positive Z-axis, which describes the orientation of the skyrmion's magnetic moments. Its functional form is as follows
\begin{equation}
\theta(r) = \pi\left[1-\exp\left(-\frac{r}{R(z)}\right)\right], \label{theta}
\end{equation}
with 
\begin{equation}
R(z) = R\sqrt{1-\frac{z^2}{h^2}}, \quad -h \leq z \leq 0, \label{Rzh}
\end{equation}
where $h$ also corresponds to the depth of the CB's Bloch point. This establishes a theoretical model that describes the 3D magnetization structures of the CB, enabling the calculation of the system's free energy.

The total free energy $E_{\text{tot}}$ of the magnetic system is obtained by integrating the magnetic energy density $\varepsilon_i$ with
\begin{equation}
E_{\text{tot}} = \iiint_V (\varepsilon_{\text{ex}}+\varepsilon_{\text{DM}}-\varepsilon_{\text{ani}}+\varepsilon_{\text{dem}}+\varepsilon_{\text{Zem}}) \, dV, \label{tot}
\end{equation}
where $V$ is the volume occupied by the CB within the magnetic material \cite{rybakovNew2016}. The total energy $E_{\text{tot}}$ includes the magnetic exchange energy $\varepsilon_{\text{ex}}=A|\nabla\mathbf{m}|^2$ with the exchange stiffness constant$A$, the chiral DMI energy $\varepsilon_{\text{DM}}=D[(\mathbf{m}\cdot\nabla)m_\mathrm{z}-m_\mathrm{z}\nabla\cdot\mathbf{m}]$ with the DMI constant $D$ \cite{fu2025}, the magnetocrystalline anisotropy energy $\varepsilon_{\text{ani}}=K(1-m_\mathrm{z}^2)$ with the magnetic anisotropy constant $K$, the demagnetization energy $\varepsilon_{\text{dem}}=-\mu_0M_\mathrm{s}\mathbf{m}\cdot\mathbf{H}_d$ with the demagnetization field $\mathbf{H}_d$, and the Zeeman energy $\varepsilon_{\text{Zem}}$ due to the external magnetic field $H_{\text{ext}}$ \cite{dong2023}. $M_s$ is the saturation magnetization intensity and $\mu_0$ is the magnetic permeability of free space. Based on the magnetization structure above, we can calculate the exchange energy of the CB as
\begin{equation}
E_{\text{ex}} = A \iiint_V \left[ \left( \frac{\partial\theta}{\partial r} \right)^2 + \left( \frac{\partial\theta}{\partial z} \right)^2 \right] dV.
\end{equation}
The perpendicular magnetic anisotropy (PMA) energy in the Z-direction via $\mathbf{|m|}=1$ is
\begin{equation}E_{\text{ani}} = -K_1 \iiint_V \cos^2\theta \, dV,
\end{equation}
where $K_1$ is the magnetic anisotropy constant of the chiral layer. The demagnetization energy is simply written as
\begin{equation}E_{\text{dem}} = \mu_0 M_s^2 \iiint_V \mathcal{N}\cos^2\theta \, dV.
\end{equation}
$\mathcal{N}$ is an effective demagnetization factor in such confined nanostructures \cite{zheng1996, dongUl2024}. In some studies of micromagnetism, the demagnetization energy, which plays a crucial role, is neglected \cite{rybakov2015, du2022}. However, it is included in our work. The Zeeman energy is
\begin{equation}E_{\text{Zem}} \approx -\mu_0 M_s H_{\text{ext}} \iiint_V \cos\theta \, dV,
\end{equation}
with an external magnetic field $H_{\text{ext}}$ along $z$ axis.
The interfacial DMI energy in the chiral magnetic (CM) material is then
\begin{equation}E_{\text{DM}} = D \iiint_V \left[ \frac{\sin(2\theta)}{2 r} + \frac{\partial\theta}{\partial r} \right] dV.
\end{equation}

This study examines the formation of a single magnetic structure within a ferromagnetic environment. In a ferromagnetic system, neighboring magnetic moments are continually aligned, resulting in a minimal angle between them. Micromagnetism provides a continuum approximation that facilitates the calculation of magnetization structures and magnetization reversal. This methodology posits that magnetization is a continuous function of position. Furthermore, it entails the derivation of relevant expressions that account for the significant contributions from exchange, magnetostatic, and anisotropy energies, and so on \cite{fidler2000}. The slow variation of magnetization intensity at the atomic lattice scale \cite{kumar2017}. Consequently, the calculation of energies leads to a simplified expression for the total energy of a CB, given by \cite{rybakov2015}
\begin{align}
E_{\text{tot}} = & -\frac{4\pi^3 R^2}{3h} A - \frac{\pi R^2 h}{3} K_1 + \frac{\pi R^2 h}{3}\mu_0 M_s^2 \nonumber \\
& + \frac{\pi^3 h R}{3} D + \pi R^2 h \mu_0 M_s H_{\text{ext}} + E_V.
\end{align}
Here, $E_V$ denotes the total energy of the uniform ferromagnetic background outside the CB volume, and it is considered a constant. By setting the derivative of the total energy to $R$ equal to zero, $\partial E_{\text{tot}}/\partial R = 0$, we obtain the following equation
\begin{equation}
\left(K_1 + \frac{4\pi^2}{h^2} A - 3\mu_0 M_s H_{\text{ext}} - \mu_0 M_s^2\right) R - \frac{\pi^2}{2} D = 0.\end{equation}
Solving this equation, we obtain the formula for the CB radius $R$, 
\begin{equation}
R(h_{\text{ext}})= \frac{\pi^2 l_{\text{D}}}{2 \left(\kappa^2 + 4\pi^2\eta^2 - 3 h_{\text{ext}} - 1 \right)}. \label{rhe}
\end{equation}
Here, the parameter $\eta = l_{\text{ex}}/h$ represents the ratio of the exchange length to the depth of the CB. The exchange length, given by $l_{\text{ex}} = \sqrt{A/(\mu_0 M_s^2)}$, characterizes the typical scale of magnetic exchange interactions. The normalized PMA is represented as $\kappa = \sqrt{K_1/(\mu_0 M_s^2)}$, and the DMI characteristic length is defined as $l_{\text{D}} = D/(\mu_0 M_s^2)$. The normalized external magnetic field is expressed as $h_{\text{ext}} = H_{\text{ext}}/M_s$. Moreover, in the absence of an external field (e.g. $H_{\text{ext}}=0$), the radius $R$ becomes $R(0)= \pi^2 l_D/[2 \left( \kappa^2 + 4\pi^2\eta^2 - 1 \right)]$.

\subsection{Annihilation mechanisms of CBs}
An external magnetic field can induce the annihilation of CBs due to the competition between micromagnetic energies in Eq. \ref{tot}. We find that magnetic fields can induce two annihilation mechanisms of the CB, depending on the rate of change of the magnetic field. The first annihilation mechanism occurs when the rate of change of the magnetic field exceeds the damping dissipation, with the energy input exceeding the damping dissipation. This prevents the system's magnetization distribution from relaxing back to its original stable state. This mechanism implies a limitation on the rate of change of the external magnetic field when manipulating CBs (e.g., with an AC field). We found that this annihilation mechanism for CBs is analogous to the droplet mode annihilation mechanism \cite{weiwe2024}. In the Weber model for a liquid droplet, two key concepts are involved: the inertial force, which disrupts the structure, and the surface tension, which maintains structural stability \cite{AMANI2024105014}. When an external magnetic field is applied, the changes in Zeeman energy lead to magnetization precession, while the damping in the magnetic system dissipates this energy. Therefore, when analyzing the droplet-like annihilation mechanism, we can compare the change in Zeeman energy, denoted as $\delta E$, to the inertial force in the droplet, and the energy dissipation, referred to as $E_{\text{dis}}$, to the surface tension.

A change in the external field $\delta H_{\text{ext}}$ induces an energy change based on the weak-field linear response, which is approximately
\begin{equation}
\delta E = \pi\mu_0 M_s R^2 h \delta H_{\text{ext}}.
\end{equation}
To calculate the dissipation energy in the magnetic system, we introduce the Rayleigh dissipation function from classical mechanics, as Gilbert did. The dissipation function for the magnetic system is
\begin{equation}
\mathcal{F} = \frac{\alpha }{2\gamma M_s} \int_V \left( \frac{\partial \mathbf{M}}{\partial t} \right)^2 d V,
\end{equation}
based on the Landau-Lifshitz-Gilbert (LLG) equation \cite{Gilbert2004}. $\gamma$ and $\alpha$ are the gyromagnetic ratio and Gilbert damping constant, respectively. From the physical meaning of the dissipation functional, the dissipation power $P$ is 
\begin{equation}
P = -2\mathcal{F} = -\frac{\alpha }{\gamma M_s} \int_V \left( \frac{\partial \mathbf{M}}{\partial t} \right)^2 d V. \label{diss}
\end{equation}
In the dynamic equilibrium process, we have an expression for the Rayleigh dissipation functional, which is
\begin{equation}
P = -\alpha \gamma \mu_0 M_s \int_V \left( \mathbf{m} \times \mathbf{H}_{\text{eff}} \right)^2 dV.
\end{equation}
In the Weber model, the dissipation energy $P$ originates from internal magnetic dissipation and damping mechanisms. Therefore, the Zeeman energy need not be included in the calculation of $P$. The magnetic field is considered the external driving force \cite{Gilbert2004}. Then, substituting the magnetization distribution into the expression for the exchange effective field, it gives
\begin{equation}
\begin{aligned}
\mathbf{H}_{\text{ex}} = & \frac{2 A}{\mu_0 M_s} \Bigg\{\Bigg[ \cos\theta( \frac{\partial^2\theta}{\partial r^2} + \frac{\partial^2\theta}{\partial z^2} ) - \sin\theta \Big[ \left( \frac{\partial\theta}{\partial r} \right)^2 \\& + \left( \frac{\partial\theta}{\partial z} \right)^2 \Big] + \frac{\cos\theta}{r} \frac{\partial\theta}{\partial r} \Bigg]\mathbf{e}_r  -\Bigg[\sin\theta \Big( \frac{\partial^2\theta}{\partial r^2} + \frac{\partial^2\theta}{\partial z^2} \Big) \\& - \cos\theta \Big[\left( \frac{\partial\theta}{\partial r} \right)^2 + \left( \frac{\partial\theta}{\partial z} \right)^2 \Big] - \frac{\sin\theta}{r} \frac{\partial\theta}{\partial r} \Bigg] \mathbf{e}_z \Bigg\}. \label{ex1}
\end{aligned}
\end{equation}
By the magnetization distribution for uniaxial PMA, we can obtain the anisotropy effective field via
\begin{equation}
\mathbf{H}_{\text{ani}} = \frac{2 K_1}{\mu_0 M_s} \cos\theta \, \mathbf{e}_z. \label{ani1}
\end{equation}
Considering only the axial contribution, the demagnetization effective field is
\begin{equation}
\mathbf{H}_{\text{dem}} = -M_s \cos\theta \, \mathbf{e}_z. \label{dem1}
\end{equation}
The DMI effective field expression gives
\begin{equation}
\mathbf{H}_{\text{DM}} =\frac{-D}{\mu_0 M_s} \left[\sin\theta \frac{\partial\theta}{\partial r} \mathbf{e}_r + \left( \cos\theta \frac{\partial\theta}{\partial r} + \frac{\sin\theta}{r} \right) \mathbf{e}_z \right]. \label{dm1}
\end{equation}
By summing Eqs. \ref{ex1}-\ref{dm1}, it gives the total effective field $\mathbf{H}_{\text{eff}}$ excluding the Zeeman contribution. Further derivation yields the expression for $\mathbf{m} \times \mathbf{H}_{\text{eff}}$, which is
\begin{equation}
\begin{aligned}
\mathbf{m} \times \mathbf{H}_{\text{eff}} =&\Big[ \frac{2 A}{\mu_0 M_s} ( \frac{\partial^2\theta}{\partial r^2} + \frac{\partial^2\theta}{\partial z^2} + \frac{1}{r} \frac{\partial\theta}{\partial r} ) + \frac{D}{\mu_0 M_s} \frac{\sin^2\theta}{r}\\ &+ \frac{1}{2} M_s \sin 2\theta - \frac{K_1}{\mu_0 M_s} \sin 2\theta \Big] \mathbf{e}_\phi. \label{mheff}
\end{aligned}
\end{equation}
We substitute Eq. \ref{mheff} into Eq. \ref{diss}, perform simplifications, and multiply by the relaxation time $\tau$ to yield the dissipated energy, which reads
\begin{equation}
\begin{aligned}
E_{\text{dis}} = &-\pi\alpha\gamma M_s \tau \Big( \frac{64\pi^2 A^2 h}{\mu_0 M_s^2 R^2} + \frac{4 K_1^2 R^2 h}{3\mu_0 M_s^2} \\&+ \frac{4\pi D^2 h}{3\mu_0 M_s^2} + \frac{\mu_0 M_s^2 R^2 h}{12} \Big).
\end{aligned} \label{Edis}
\end{equation}
Here, the negative sign indicates energy dissipation. The relaxation time $\tau\sim$ ns can be obtained by micromagnetic simulations based on the LLG equation.  Analogous to the liquid droplet, we define the dimensionless magnetic Weber number
\begin{equation}
We = \left| \frac{\delta E}{E_{\text{dis}}} \right|.
\end{equation}
Therefore, the magnetic Weber number can be written as,
\begin{equation}
We=\frac{12\mu_0^2 M_s^2 R^4 |\delta H_{\text{ext}}|}{\alpha\gamma\tau \left( 768\pi^2 A^2 + 16K_1^2 R^4 + 16\pi D^2 R^2 + \mu_0^2 M_s^4 R^4 \right)}, \label{we}
\end{equation}
for CBs. When $We\geqslant 1$, the magnetic bobber annihilates, when $We < 1$, the bobber remains stable under the field change \cite{liEncyc2008}. Setting $We = 1$ yields the relation between the maximum magnetic field step $\delta H_c$ and the CB radius $R$, namely 
\begin{equation}\delta B_c=\mu_0\delta H_c \approx C/R^2
, \label{bc} \end{equation}
where $C \approx 24.9$ T$\cdot$nm$^2$ is a constant incorporating magnetic parameters. According to Eq. \ref{bc}, larger CB radii $R$ tolerate more minor changes in the external field $\delta H_{\text{ext}}$. When the external magnetic field changes $\delta H_{\text{ext}}$ is greater than $\delta B_c$, the CBs undergo a fast annihilation process.

The second annihilation mechanism arises from the pinning effect associated with the CB's Bloch point. Due to this pinning effect, the depth $h$ of the Bloch point remains almost constant even as the external magnetic field changes. The Bloch point forms at the interface between two layers with different PMA and DMI. Variations in PMA and DMI across this interface create a localized energy barrier that hinders the motion and transformation of the topological defect, leading to pinning of the Bloch point \cite{charilaou2020}.

The Bloch point is a vacuum point where magnetization is undefined, surrounded by magnetic moments. To describe the role of the localized energy barrier, we consider the Bloch point and the magnetization distribution within a sufficiently small surrounding region as a whole \cite{lang2023}. As the external field gradually increases, the energy of the Bloch point increases until it overcomes the localized energy barrier, leading to the annihilation of the Bloch point and, consequently, the annihilation of the CB. To quantify this energy barrier, we describe the 3D magnetization distribution around the Bloch point as \cite{tejo2023}
\begin{equation}
\mathbf{m}(r,\theta,\phi) = (\sin\theta\cos\phi, \sin\theta\sin\phi, \cos\theta), 
\end{equation}
with $0 \leq r \leq R_B$. $R_B$ is the radius of a sufficiently small sphere centered on the Bloch point. $R_B$ is generally on the nanometer scale and depends on the specific material. We verified the correctness of its value through micromagnetic simulation with $R_B = 1 \sim 5$ nm. The localized energy barrier primarily arises from the differences in PMA and DMI between the two layers. Considering the magnetization distribution and computing the difference between the anisotropy and DMI energies in the two layers, we have the energy barrier, 
\begin{equation}
E_{\text{bar}} = -\frac{8\pi R_B^3}{9} \Big(K_2 - K_1\Big) + \frac{2\pi R_B^2}{3} D.
\end{equation}
The energy required to overcome the barrier $E_{\text{bar}}$ results from the Zeeman energy of the external field. Substituting the magnetization distribution into the Zeeman energy, it gives
\begin{equation}
E_{\text{Zem}} = -\frac{2}{3} \pi \mu_0 M_s R_B^3 H_{\text{ext}}.
\end{equation}
By $E_{\text{Zem}}(H_{\text{th}}) = E_{\text{bar}}$, it gives the threshold magnetic field $H_{\text{th}}$ at which the Bloch point annihilates, causing the CB to annihilate, which is
\begin{equation}
B_{\text{th}} = \mu_0 H_{\text{th}} = \frac{4}{3 M_s} \Big(K_2 - K_1\Big) - \frac{D}{M_s R_B}. \label{Bth}
\end{equation}

\subsection{Micromagnetic simulations}
We conduct a validation of our proposed theoretical model for CBs through the use of the Mumax3 micromagnetic simulation \cite{joos2023}. In our simulations, we denote the anisotropy constant of the upper CM material as $K_1$, while the strong PMA constant of the lower material, $K_2$, is varied. The chiral bilayer is structured within a magnetic material with overall dimensions of $150 \times 150 \times 25$ nm and a simulation cell size of $1 \times 1 \times 1$ nm, as shown in Fig.\ref{fig1}(a). This nanoconfinement structure comprises two layers designed to create CBs: a lower layer of standard magnetic material with a thickness of 9 nm and an upper chiral layer featuring interfacial DMI with a thickness of 19 nm.

The magnetic parameters in both layers differ only in terms of the DMI constant $D$ and the PMA constant $K$. In contrast, all other magnetic parameters remain the same. We employed multiple algorithms and ensured that sufficiently long relaxation times were applied to guarantee the reliability of our results. The material parameters used in the simulations are as follows: the saturation magnetization is set to $M_s = 500 \, \text{KA/m}$ and the exchange constant is $A = 7 \, \text{pJ/m}$. The Gilbert damping coefficient $\alpha$ is initially set to 0.5 during magnetization setup and throughout the relaxation process to achieve the minimum energy state, thereby achieving a stable magnetic structure. During the external field excitation phases, $\alpha$ is reduced to 0.01. The upper chiral layer has an interfacial DMI constant of $D = 2.5 \, \text{mJ/m}^2$ and a PMA constant of $K_1 = 500 \, \text{KJ/m}^3$. The perpendicular anisotropy constant for the lower layer is initially set to $K_2 = 1 \, \text{MJ/m}^3$, and this constant is varied subsequently to study the effects of differences in anisotropy.

\section{Results and Discussion}
\begin{figure} 
\setlength{\abovecaptionskip}{0.1cm}
\includegraphics[width=0.48\textwidth, angle= 0]{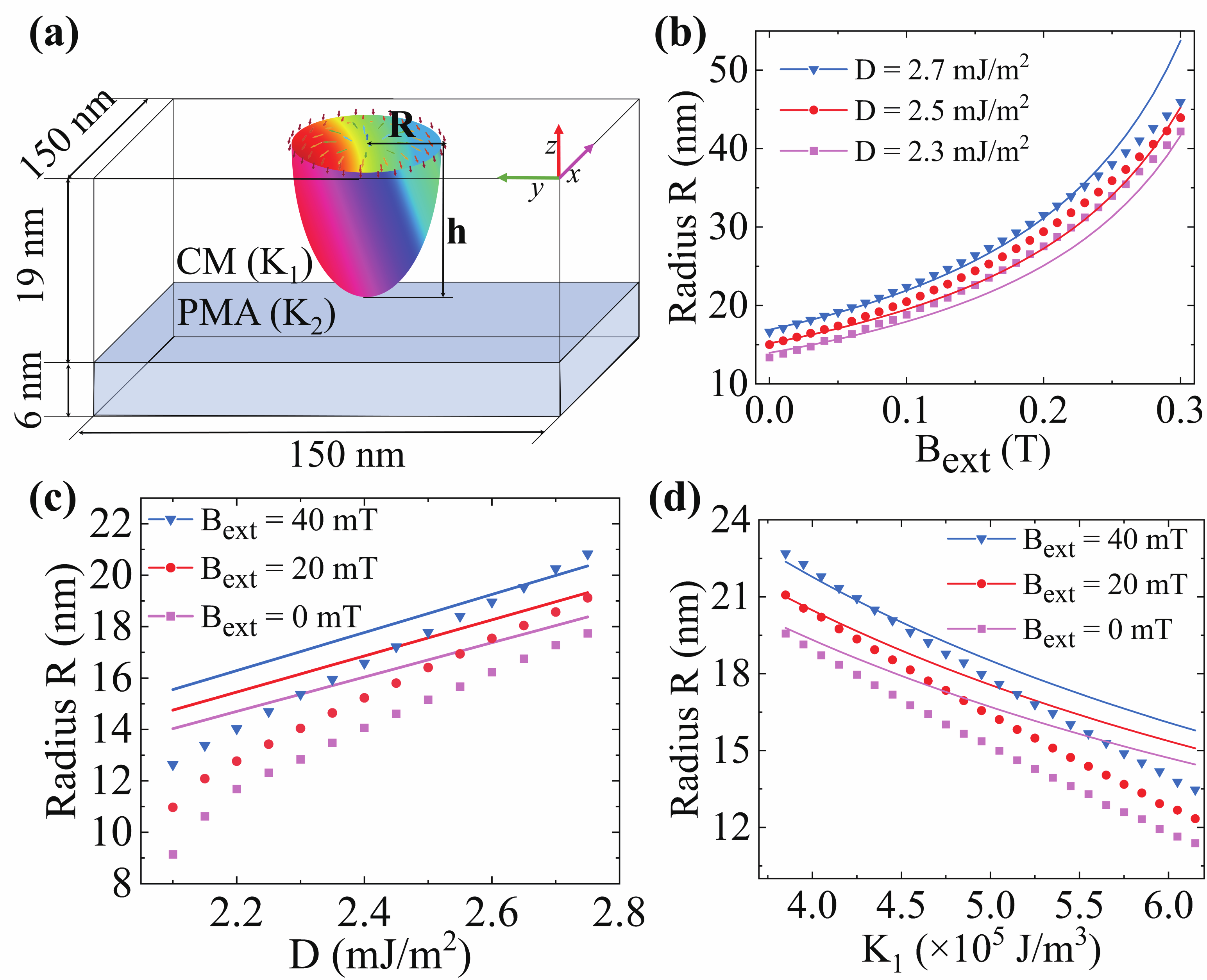}
\caption{(a) Bilayer magnetic nanostructures and the generated single CB structure with the radius $R$ and the depth $h$. (b) Variation of radius $R$ of CBs with external magnetic field B$_{\text{ext}}$ for the different DMI $D$. (c) The radius $R$ of CBs vs. $D$ with different B$_{\text{ext}}$. (d) The radius $R$ of CBs vs. PMA $K_1$ with the different B$_{\text{ext}}$.} \label{fig1}
\end{figure}
Fig. \ref{fig1}(a) illustrates the bilayer nanostructure, which consists of an upper chiral magnetic layer with DMI and a lower ferromagnetic (FM) layer dominated by the strong PMA material. It shows the achieved 3D topological structure of a CB, highlighting its key parameters, the radius $R$ and the depth $h$ \cite{gong2021}, as shown in our theoretical Eqs. \ref{theta} and \ref{Rzh}. This theoretical model describes the magnetization of the CB as it undergoes a continuous rotation from the Bloch point (located at $z = -h$) to the skyrmionic texture, as the diameter decreases from the surface (at $z = 0$). The exponential decay ansatz for the polar angle $\theta(r, z)$ effectively captures the 3D morphology, allowing for analytical energy minimization.

Fig. \ref{fig1}(b)-(d) rigorously validates the theoretical size formula for $R$ (see Eq. \ref{rhe}) against micromagnetic simulations across key parameters. At a fixed value of $K_2 = 1$ MJ/m$^3$, the radius $R$ of CBs steadily increases as the magnetic field B$_{\text{ext}}$ rises. There is a small discrepancy between the theoretical predictions (solid lines) and the results (symbols) from micromagnetic simulations. The primary reason for this difference stems from our theoretical analysis. To obtain an analytical solution, we exclude the energy contribution from Bloch points and simplify the calculation of the demagnetization energy, as seen in Fig. \ref{fig1}(b). Accurately calculating the contribution of the demagnetizing field enables more precise results. In Fig. \ref{fig1}(c), $R$ increases with increasing DMI, as a larger DMI results in a larger spiral cycle. At $B_{ext} = 0 $ T, $ R(D)$ scales according to $ R(0) = \left( \kappa^2 + 4\pi^2\eta^2 - 1 \right)^{-1} $, matching the simulations within a $5\%$ error margin. Increasing $K_1$ compresses $R$ due to the enhanced PMA. The results of our simulations confirm the theoretical predictions regarding the inverse dependence of $R$ (e.g., $\kappa$ in Eq. \ref{rhe}), as illustrated in Figure \ref{fig1}(d). Our theoretical findings indicate that nanoscale sizes of spin structures can be achieved by optimizing size-field tunability.

We present a theoretical model that describes the size formula for CBs. This model establishes a functional relationship, $R$($B_{ext}$, $A$, $D$, $K_1$) for 3D CBs, which was previously unavailable. This relationship effectively addresses the interplay between magnetic exchange, chiral DMI, PMA, and Zeeman energies. The remarkable agreement between micromagnetic simulations and our analytical results confirms the validity of our approach, despite the complexity of three-dimensional topological structures. Fig. \ref{fig1} demonstrates the consistency of trends between our analytical predictions and the simulation results, thereby elucidating how key parameters govern the size of the CB. The models we derived establish foundational principles for utilizing CBs in three-dimensional spintronic devices in experimental settings. 

\begin{figure} 
\setlength{\abovecaptionskip}{0.1cm}
\includegraphics[width=0.48\textwidth,angle= 0]{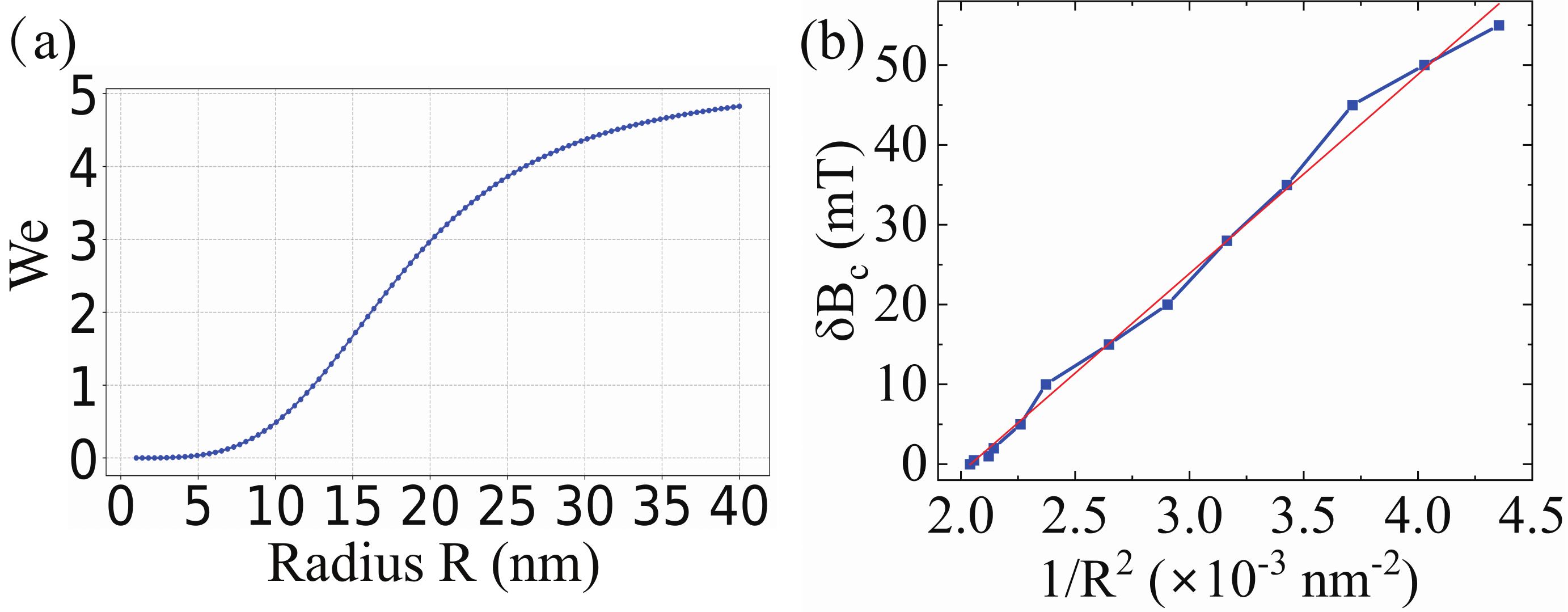}
\caption{(a) The magnetic Weber number $We$ as the function of the CB radius $R$. (b) The maximum magnetic field step $\delta B_c$ by the simulations as a function of $1/R^2$ of CBs.} \label{fig2}
\end{figure}
Fig. \ref{fig2}(a) quantifies the first annihilation mechanism using the magnetic Weber number $We$, showing a nonlinear relationship with the critical bubble radius $R$, which aligns with the proposed droplet-like model. The monotonic increase indicates a clear instability in larger CBs under a rapid magnetic-field changes. This suggests that a larger $R$ can enhance the injection of Zeeman energy during fluctuations in the magnetic field. Additionally, the damping dissipation efficiency decreases as $E_{\text{dis}} \propto R^{-2}$ approximately, tuned by the exchange and the DMI as described in Eq. \ref{Edis}. This model resembles classical droplet models and emphasizes the unique 3D energy topology associated with critical bubbles \cite{DUAN01072003}.  

Fig. \ref{fig2}(b) validates the universal scaling law for dynamic annihilation thresholds. The linear correlation $\delta B_c \propto 1/R^2$ (red theory vs. blue simulations) confirms the theoretical framework, $\delta B_c = C/R^2$ with $C \approx 24.9$ T$\cdot$nm$^2$. This $1/R^2$ dependence establishes a quantitative criterion for manipulating collective behavior, enabling the predictive design of AC-field controls for nonvolatile CB-based devices. The superlinear relationship $We \propto R^3$ (see Fig. \ref{fig2}(a)) reveals unique 3D CB textures, which act as damping models for topological spintronics. By combining Figs. \ref{fig1} and \ref{fig2}, we observe dual stability regimes. It is shown that small CBs are dominated by Bloch-point depinning, while large CBs are governed by droplet-like annihilation, as illustrated in Fig. \ref{fig2}. The exact match between theory and simulation demonstrates that the analytical droplet-like model can replace costly dynamic simulations for predicting the stability of 3D CBs. Fig. \ref{fig2} conclusively establishes the $1/R^2$ scaling of dynamic annihilation thresholds for 3D magnetic quasiparticles. Additionally, the nonlinearity of $ We$ highlights fundamental differences between CBs and skyrmions, demonstrating the role of dimensionality in topological stability \cite{gaD2022}. These results provide a quantitative foundation for manipulating CBs in spin-based topological devices, where magnetic field-step tolerance is critical for operational reliability \cite{bald2025}.

\begin{figure} 
\setlength{\abovecaptionskip}{0.1cm}
\includegraphics[width=0.48\textwidth,angle= 0]{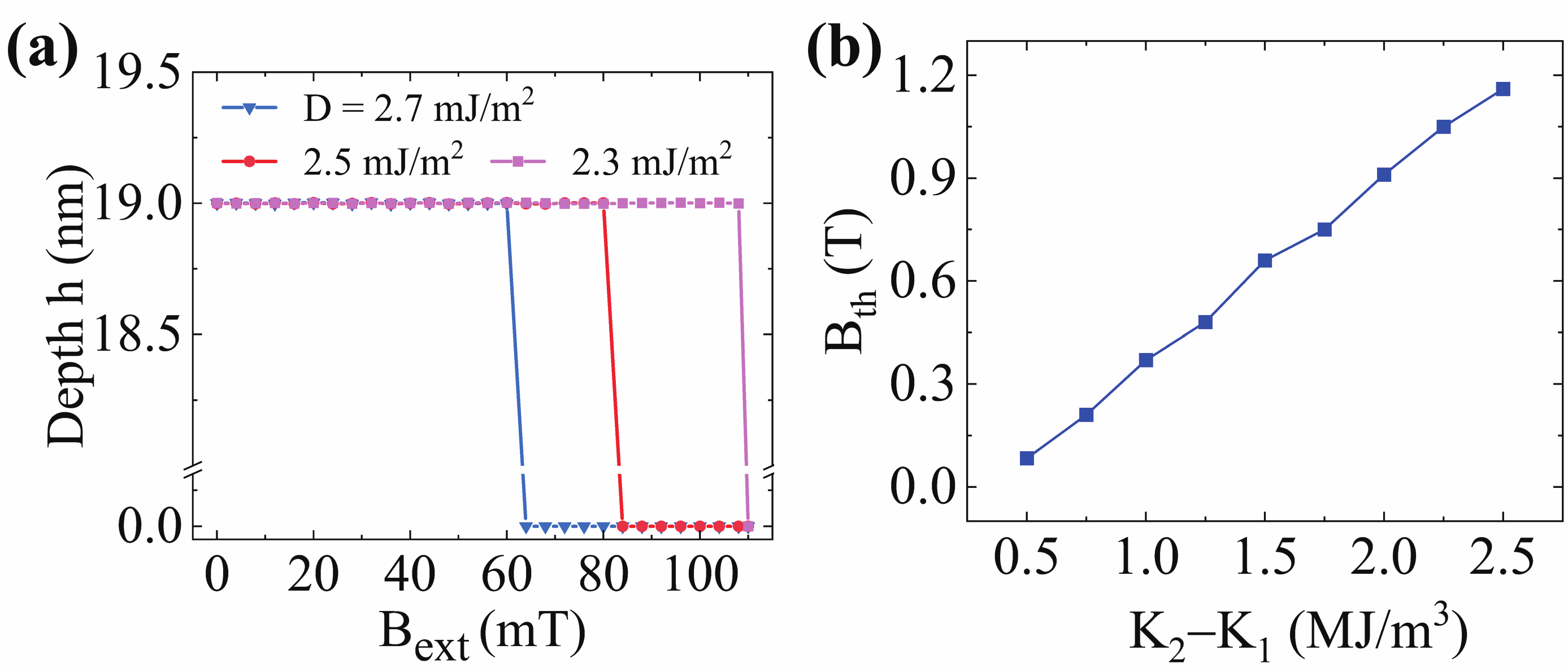}
\caption{(a) The depth $h$ of the CB as a function of external fields B$_{\text{ext}}$. (b) Threshold magnetic field B$_{\text{th}}$ for the first magnetic bobber annihilation dependent on $K_2-K_1$. } \label{fig3}
\end{figure}

Fig. \ref{fig3}(a) illustrates the critical annihilation transition of CBs as a function of the depth parameter $h$. The step-function collapse observed as $h$ approaches $0$ across different DMI values supports the Bloch-point depinning mechanism outlined in Eq. \ref{Bth}. The plateau in $h$ before annihilation confirms that the Bloch point is pinned, preventing depth modulation until certain critical energy thresholds are surpassed. Before annihilation occurs, the depth $h$ of a CB remains constant (for instance, $h = 19$ nm). The critical thresholds required for annihilation increase in magnetic fields at lower D values, consistent with the theoretical scaling of $B_{\text{th}}$ presented in Eq. \ref{Bth}. The depth $h$ of Bloch points depends on the thickness of the upper chiral magnetic layer (e.g., Fig. 1(a)), as well as the anisotropy difference $K_2-K_1$ between the upper and lower layers (chiral and PMA materials), as expressed in Eq. \ref{Bth}. Once CBs form, $h$ remains stable.  An external field must overcome the pinning energy $E_{\text{bar}}$ at the Bloch point to annihilate CBs. The value of $h$ undergoes a significant change. The evolution of CBs with film thickness has been experimentally observed \cite{ranBobber2021, ahmedChiral2018}. A simple theoretical pinning model of the Bloch point is shown \cite{gong2021}. Fig. \ref{fig3}(b), which demonstrates a universal linear relationship with a slope of 0.45 T/(MJ/m$^3$) by the simulations, which aligns closely with the theoretical expression $B_{\text{th}} \propto (K_2 - K_1) \propto -D$. These findings provide strong validation of Eq. \ref{Bth} as the governing law for quasi-static annihilation, thereby distinguishing it from skyrmion-collapse models \cite{rohart2016}. The linear relationship between $B_{\text{th}}$ and $\Delta K$ facilitates predictive material design. For instance, achieving stability thresholds of 0.5 T requires $\Delta K > 1.1$ MJ/m$^3$.  

The consistency between the simulated $B_{\text{th}}$ values and our analytic predictions indicates that our proposed magnetic continuum models are effective for analyzing 3D textures at both atomic and device scales. Fig. \ref{fig3} shows that the difference in interfacial anisotropy is the primary factor determining the annihilation thresholds of the CB states, while the DMI plays a secondary stabilizing role. The linear relationship observed in the $B_{\text{th}}$ values provides a universal approach for modeling 3D CB states. This work highlights that CB states, as nontrivial quasiparticles, exhibit distinct collapse phenomena when compared to skyrmions or bubbles \cite{Cai2012}.

\begin{figure} 
\setlength{\abovecaptionskip}{0.1cm}
\includegraphics[width=0.48\textwidth,angle= 0]{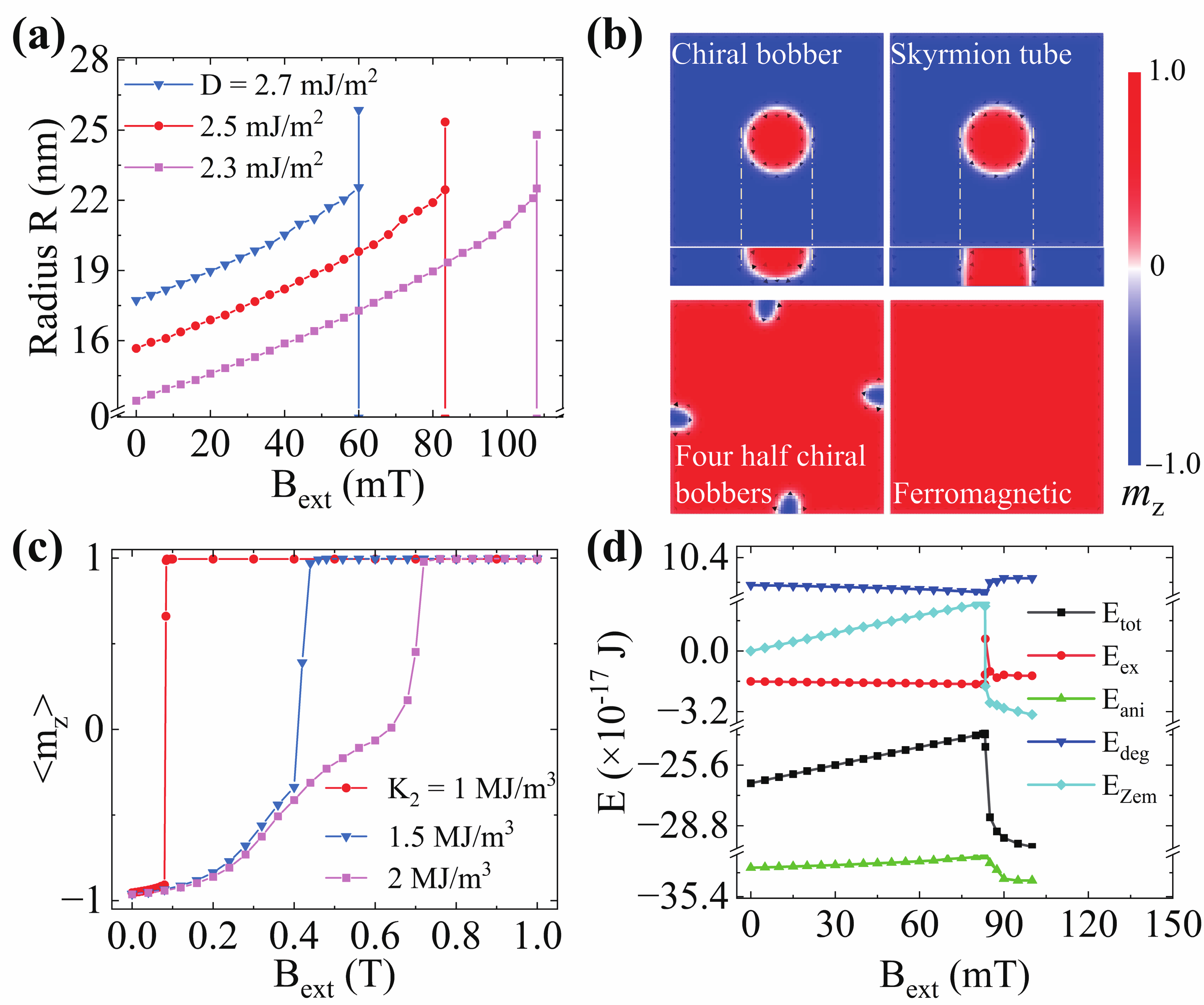}
\caption{(a) The radius $R$ and the second annihilation of CBs with external magnetic field $B$ for different DMI $D$. (b) The second annihilation into CBs from a CB to an ST, an HCB, and an FM state. (c) Average magnetization $m_z$ as the magnetic field changes for different $K_2$. (d) The total magnetic energy $E_{\text{tot}}$, exchange energy $E_{\text{ex}}$, the PMA energy $E_{\text{ani}}$ and the demagnetization energy $E_{\text{deg}}$ as a function of external magnetic fields. } \label{fig4}
\end{figure}

\begin{figure} 
\setlength{\abovecaptionskip}{0.1cm}
\includegraphics[width=0.48\textwidth,angle= 0]{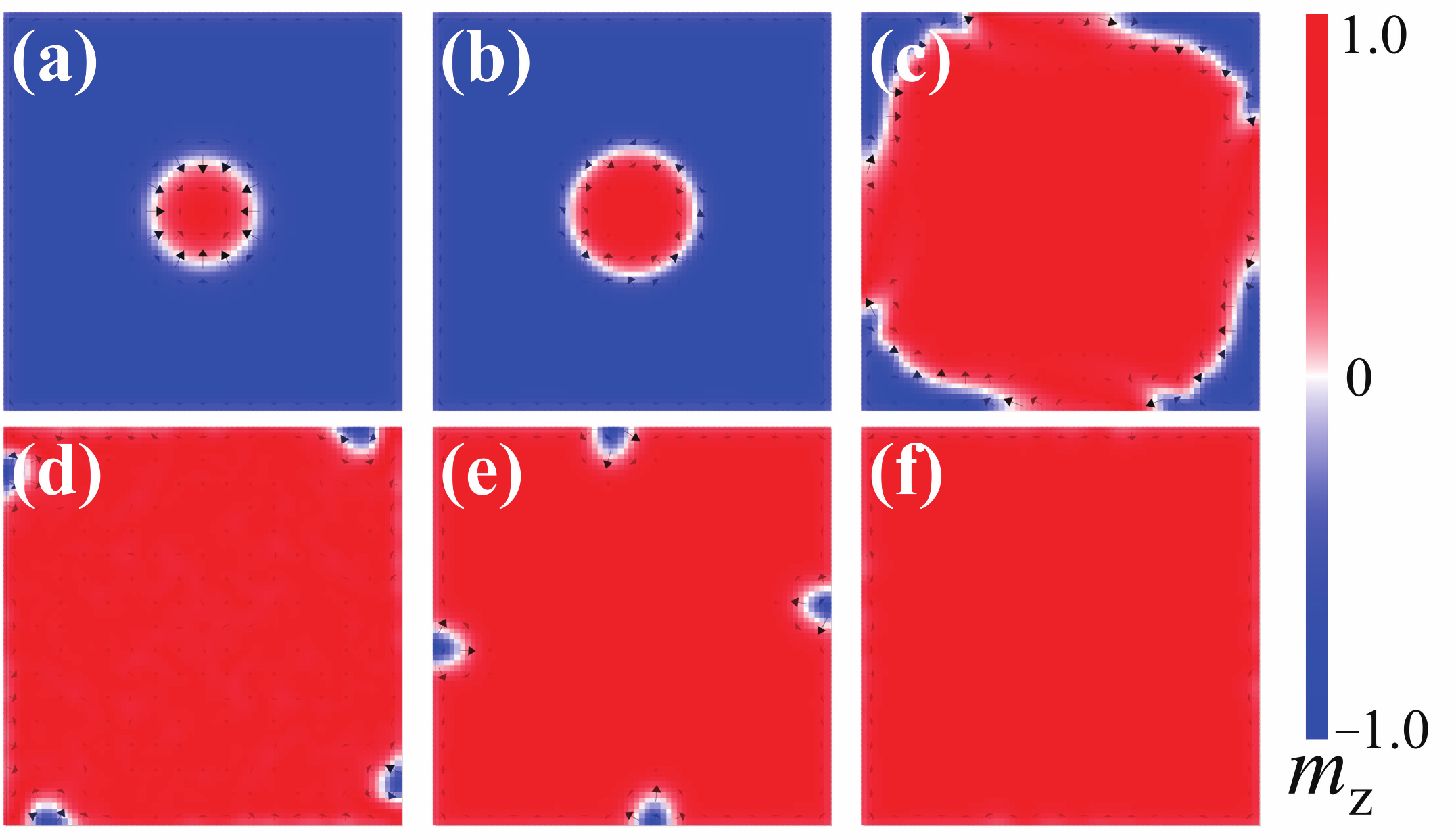}
\caption{The snapshot of the annihilation process of a CB from a CB to an ST, an HCB, and an FM state.} \label{fig5}
\end{figure}

Fig. \ref{fig4}(a) illustrates the critical role of DMI strength $D$, in dictating the second annihilation mechanism. As indicated by Eq. \ref{Bth}, a lower value of $D$ delays the annihilation process to higher magnetic fields. This suggests that a larger chiral $D$ facilitates the expulsion of Bloch points from CM nanostructures. This behavior contrasts significantly with that of 2D skyrmions. Fig. \ref{fig4}(b) introduces a novel annihilation pathway for CBs: the sequence is CB $\to$ skyrmion tube (ST) $\to$ four half-CBs (HCBs) $\to$ FM state. The slow annihilation process corresponds to a small change of the external magnetic field $\delta B_c$. This pathway demonstrates that the transition from CB to ST preserves a topological charge of $Q = 1$. However, the subsequent decay into HCBs divides $Q$ into four $Q = 1/4$ topological textures. We have discovered that a similar phenomenon is also observed in skyrmions, but the transformation process differs significantly between 2D and 3D systems \cite{linsz2015}. A more detailed annihilation process is shown in Fig. \ref{fig5}, and a detailed demonstration of this ultrafast annihilation process in Video 1 of the Supplementary Material \cite{supply}. Our study indicates that the annihilation process for 3D topological structures is significantly more complex than in 2D due to the increased dimensionality, particularly in the presence of Bloch points. The pinning of Bloch points endows 3D CBs with a unique annihilation mechanism. The mechanism behind the phase transition of 3D magnetic topological structures requires further investigation.

Fig. \ref{fig4}(c) demonstrates how the uniaxial anisotropy $K_2$ in the low magnetic layer influences pathway selection while keeping $K_1 = 500 \, \text{KJ/m}^3$ constant. When $K_2$ is low, we observe single-step $m_z$ switching, indicating a direct transition from CB to FM annihilation. Conversely, a high $K_2$ leads to three-step transitions that correlate with the topological changes illustrated in Fig. \ref{fig4}(b). Fig. \ref{fig4}(d) presents the variation in the magnetic field as a function of the micromagnetic energy, highlighting a significant energy change during the annihilation of the CB, driven by magnetic fields. Fig. \ref{fig4} and \ref{fig5} provide a clear picture of annihilation, characterized by multi-stage fragmentation mediated by STs. It reveals a unique annihilation pathway influenced by chirality that distinguishes 3D magnetic spin textures from 2D skyrmions.

\section{Conclusion}  
We have resolved the size-tuning and novel annihilation mechanisms of CBs through integrated analytical modeling and micromagnetic simulations. Our derived expression for the CB radius $R$ quantitatively captures the competition among Zeeman energy, DMI, and anisotropy, and is validated across a range of field strengths and material parameters. We identify dual annihilation pathways: Dynamic annihilation under rapid field changes follows a droplet-like instability with magnetic Weber number $We$, indicating heightened vulnerability in larger CBs. Quasi-static annihilation arises from Bloch-point depinning at interfaces, where B$_{\text{th}}$ scales linearly with anisotropy gradients and inversely with DMI strength.  

We have identified a unique fragmentation pathway: CBs transition to STs, which then break down into four HCBs, ultimately leading to FM states. This pathway is absent in 2D systems and exhibits markedly different behavior compared to that of skyrmions. During this process, the topological charge transitions from integer to fractional values, highlighting the influence of dimensionality on the dynamics. Our research establishes key design principles for the stability of circular bubbles. Small circular bubbles are primarily affected by Bloch-point pinning, while larger ones tend to collapse in a droplet-like manner. These insights facilitate predictive control of circular bubbles in applications such as racetrack memories and neuromorphic computing, thereby advancing the field of 3D spintronics.

\begin{acknowledgments}
This work is supported by the National Natural Science Foundation of China (Grant Nos. 12204497 and 12374079) and funded by Science and Technology Program of Xuzhou (KC25001).
\end{acknowledgments}


\nocite{}
\bibliography{ref}
\end{document}